\documentclass[preprint,a4paper,twocolumn,12pt]{aastex63}
\usepackage{fancyhdr}
\begin{document}

\title{Deep-learning based measurement of planetary radial velocities in the presence of stellar variability}
\correspondingauthor{Alexander Wise}
\email{aw@psu.edu}

\correspondingauthor{Virisha Timmaraju}
\email{virisha.timmaraju@jpl.nasa.gov}
\author[0000-0001-8126-4259]{Ian Colwell}
\affiliation{NASA Jet Propulsion Laboratory, California Institute of Technology, Pasadena, CA 91011
}

\author[0000-0001-5688-646X]{Virisha Timmaraju}
\affiliation{NASA Jet Propulsion Laboratory, California Institute of Technology, Pasadena, CA 91011
}

\author[0009-0004-0856-0018]{Hamsa Shwetha Venkataram}
\affiliation{NASA Jet Propulsion Laboratory, California Institute of Technology, Pasadena, CA 91011
}

\author[0000-0002-5013-5769]{Alexander Wise}
\affiliation{Department of Astronomy and Astrophysics, Penn State University, State College, PA 16802}

\begin{abstract}
    We present a deep-learning based approach for measuring small planetary radial velocities in the presence of stellar variability. We use neural networks to reduce stellar RV jitter in three years of HARPS-N sun-as-a-star spectra. We develop and compare dimensionality-reduction and data splitting methods, as well as various neural network architectures including single line CNNs, an ensemble of single line CNNs, and a multi-line CNN. We inject planet-like RVs into the spectra and use the network to recover them. We find that the multi-line CNN is able to recover planets with 0.2 m/s semi-amplitude, 50 day period, with 8.8\% error in the amplitude and 0.7\% in the period. This approach shows promise for mitigating stellar RV variability and enabling the detection of small planetary RVs with unprecedented precision.
\end{abstract}

\section{Background} \label{sec:backg}

Stellar noise is one of the most problematic noise sources in Extreme Precision Radial Velocity (EPRV) measurements \citep[e.g.][]{fischer16,nas18}. Broadly speaking, stellar noise in EPRV measurements can be mitigated in either the time domain (FF’ or a GP) or the wavelength domain (modeling the flux or some function of the flux). Time domain methods use activity proxies such as photometry \citep[original FF' method,][]{aigrain12}, unsigned magnetic flux \citep{haywood22}, or activity indicators derived from the spectra and often use GPs to model the RVs and stellar noise simultaneously \citep{rajpaul15,jones17,gilbertson20}. Wavelength-domain methods can be used to generate activity indicators using spectra \cite[e.g.][]{wise19, ning19, siegel22}, or to measure or correct RVs via clever modeling of the spectra or cross-correlation function (CCF) \citep[e.g.][]{dumusque18,cretignier20a,rajpaul20,colliercameron21,deBeurs22,zhao22,cretignier22}.

Here we present a new approach for modeling the spectra in the wavelength domain where we use Deep Learning (DL) based neural networks to measure small planetary RVs in the presence of stellar noise. \cite{deBeurs22} show that the CCF has enough information for a neural network to reduce stellar RV jitter in three years of HARPS-N sun-as-a-star spectra down from 1.47 m/s to 0.78 m/s. To push towards more precise corrections, we will leverage the differences between spectral lines’ responses to stellar activity \citep[e.g.][]{wise19} by using spectra instead of CCFs as our dataset. The unprecedented Signal-to-Noise Ratio (SNR) and cadence of sun-as-a-star spectra allow us to evaluate the effectiveness and limitations of neural networks at separating stellar and planet-induced RVs in the wavelength domain at sub-m/s precision, and determine their applicability to the EPRV community’s goal of mitigating stellar RV variability.

\section{Motivation and Objectives} \label{sec:objs}

Neural networks are able to capture the complex non-linear relationships between the spectral signatures of different kinds of stellar variability and RVs that are challenging for traditional EPRV approaches to capture. Hence our choice to develop DL pipelines and to explore their applicability to separating planetary and stellar activity RVs stems from a perceived potential push the limits of stellar activity correction in EPRV measurements. 

The primary objective of this paper is to describe the deep learning-based pipeline we developed to measure planetary RVs in spectra. 

In Section \ref{sec:data_sources} we describe the data sources, Section \ref{sec:data_prep} covers data preparation and preprocessing steps, Section \ref{sec:ml} dives into the different machine learning approaches we developed, Section \ref{sec:results} looks at the results of the machine learning approaches, Section \ref{sec:conclusiondiscussion} summarizes our work and discusses future areas for improvement.

\section{Data Sources} \label{sec:data_sources}

The primary dataset that we use for this work is the HARPS-N Solar Spectra (HARPS-n) a high-resolution RV spectrograph with continuous wavelength coverage from ~380 to 690 nm located on the island of La Palma, Canary Islands, Spain. In September 2020, HARPS-N released their first public dataset: 34550 disk integrated spectra of the sun taken from 2015 to 2018. This dataset includes significant numbers of sunspots, faculae, and plage, and a higher-than-average level of convective blueshift variation due to the solar magnetic cycle maximum and minimum that occurred in 2014 and 2019 respectively. The solar spectra have 5-minute integration times to average down solar p-mode oscillations, and have average SNRs of approximately 350. The data are obtained from the University of Geneva Data \& Analysis Center for Exoplanets website (dace.unige.ch).

This dataset has or is expected to have stellar activity signals from pulsations, granulation, and magnetic activity in the form of spots, faculae, plage, and convective blueshift variations.



\section{Data Preparation} \label{sec:data_prep}
To evaluate the ability of DL-based neural networks to separate stellar and planetary RVs, we utilize end-to-end injection and recovery of planet-like Doppler signals. We start with extracted 2D spectra, and inject the planetary signals by adding them to the heliocentric frame correction RVs as soon as possible. This way when we interpolate the spectra onto a common wavelength grid, the planet-like signals are already included. The 2D spectra are normalized before interpolating onto a common grid. The steps are summarized below.

\subsection{Pre-processing of HARPS Spectra}\label{sec:data_prep_harps}
We begin our data-processing with extracted 2-D spectra (in HARPS filenames, ‘.e2ds’ or ‘.s2d’ files). The first steps we perform are order-by-order normalization, RV shifting, and interpolation onto a common wavelength grid. The spectra are normalized using a Julia implementation of the RASSINE method \citep{cretignier20b} in the public NeidSolarScripts.jl package. Normalization is an important step to make sure the ML approach does not simply focus on the highest-flux pixels because they have the highest variations (as is true for photon noise). But it has the drawback of losing information about the amount of percent variation expected in each pixel due to photon noise. However, this information loss is low due to the extremely high SNR of the solar spectra.

After normalization, we Doppler shift the normalized spectra (henceforth referred to simply as spectra) to remove barycentric motion, and interpolate all of the spectra onto a single wavelength grid (from one observation) using a sinc kernel for interpolation, which prevents the introduction of noise due to intra-pixel sensitivity. Interpolating onto the original pixel grid allows us to preserve the spectrum’s pixel sizes, limiting interpolation uncertainty that would be caused by changing the wavelength grid spacing. On the HARPS-ACB data, the RVs measured by the mask CCF method before and after sinc interpolation have an RMSE of 11 cm/s introduced by the interpolation, significantly lower than a linear or cubic spline interpolation, but equal in RMSE to sampling from a gaussian process with a matern 5/2 kernel in this before-and-after test. Thanks to the identical wavelength grid spacing in HARPS and HARPS-N, we can interpolate spectra from both of these instruments onto a common wavelength grid without changing bin sizes. After interpolation, we combine the orders into a 1-D spectrum by removing the overlapping order edges, throwing away overlapping pixels that are furthest from an order center as these have the lowest SNR. Finally, we remove data below 420 nm due to their relatively high blending and photon noise and above 690 nm due to oxygen telluric lines.

During the Doppler shifting step, to remove barycentric motion, we inject a planetary signal. These planet-induced RVs become the target values for the neural network to predict. In the case of solar spectra, the barycentric corrections remove the RVs due to all significant celestial bodies, so we can be sure that the injected planetary signals are the only center-of-mass RVs in the spectra, and remaining RV signals are due to stellar jitter. This makes for an ideal test dataset for this approach, which proposes to use wavelength-domain information to distinguish between stellar and planet-induced RVs.

\subsection{Injecting Planet-like RVs into Spectra}\label{sec:data_prep_rvs}
Building off the preprocessing steps applied to the HARPS spectra, we create multiple datasets where the injected RV signals are selected to support distinct steps necessary to develop and evaluate the neural network.

\subsubsection{Random RVs}\label{sec:data_prep_rvs_random}
To facilitate training the neural network we developed datasets composed of random uniform RVs. Given a single solar spectrum, a random RV was selected between the range of [-1000, 1000] m/s, and the spectrum was shifted based on that RV. This is the same RV injection described in the Doppler shifting step in Section \ref{sec:data_prep_harps}. This process was performed 60 times for each solar spectrum in HARPS-N, resulting in a dataset composed of 2073000 spectra. We refer to this dataset as the Random Uniform 1000 m/s RV Spectra. 

The Random Uniform 1000 m/s RV Spectra are intended to prevent the neural network from overfitting by exposing the network to same pattern of spectral noise (by using the same spectrum) multiple times, but each time with a different injected planetary RV target. Creating a dataset where each spectrum is considered in isolation, when injecting the planet-like RV, is possible because the neural network is not trained with data structured in a temporal manner.

\subsubsection{Planet-like RVs}\label{sec:data_prep_rvs_planet}
\begin{figure}
\includegraphics[width=\columnwidth]{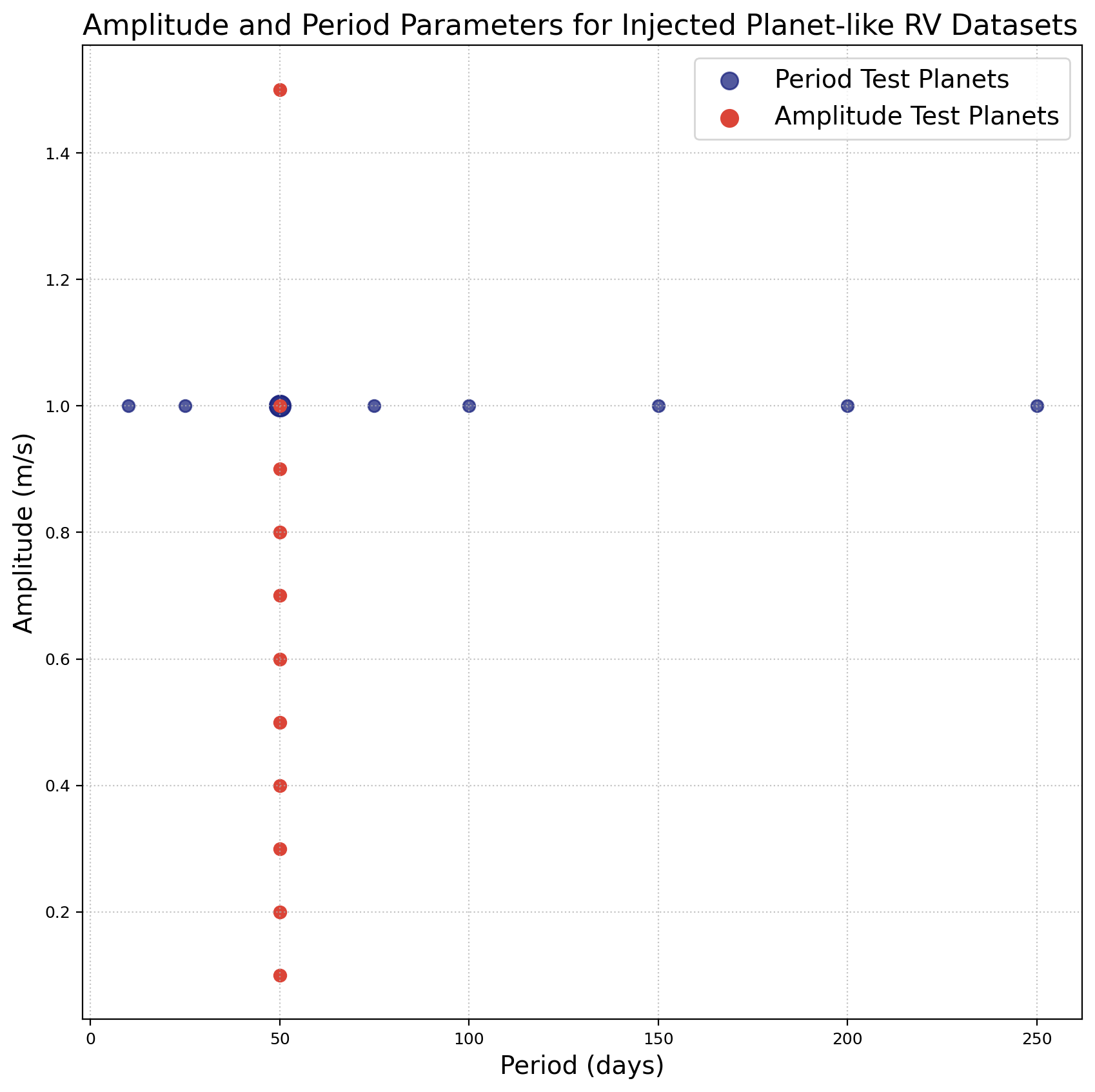}
\caption{Full planet-like RV test cases used for evaluating the neural network's ability to recover planet-like RVs. Purple points correspond to planets testing shorter and longer period orbits (keeping amplitude fixed at 1 m/s), while red points correspond to planets testing for amplitude (keeping period fixed at 50 days). }
\label{fig:planets_amp_per_tests}
\end{figure}

The overarching goal of this work is to develop a neural network that can accurately output activity-corrected RVs given an input spectrum. To this end, we inject sinusoidal planet-like RV signals into the HARPS-N spectra to create a series of planet test cases. The planet test cases are aimed at determining a trained neural networks sensitivity of two parameters, RV amplitude and period. The set of planets focused on testing amplitude have a fixed period of 50 days while the amplitude ranges from 0.1 m/s to 1.0 m/s, in increments of 0.1 m/s. The set of planets focused on testing period have a fixed amplitude of 1.0 m/s while the period ranges from 10 days to 250 days, in varying increments of 15, 25, and 50 days. A total of 20 planet-like RV datasets were created, 10 for each of the period and amplitude test cases (Figure \ref{fig:planets_amp_per_tests}).

\subsection{Selecting Spectral Lines}\label{sec:data_prep_mask}
We utilize a visible-light spectral line list of 4570 lines to extract lines from the HARPS-N spectra, where the window around each line spans 15 pixels. While neural networks are capable of handling the full spectrum as input, there are some disadvantages to this approach. Input dimensionality will be much larger, which in most cases, will correspond to a larger number of trainable parameters and increase the time to train the network. Our primary goal was to determine the applicability of neural networks in measuring RVs, as such, enabling reasonable training times to allow quick iterations was prioritized.

\subsection{Train, Validation, and Test Data Splits}\label{sec:data_prep_splits}
Data are split into training, validation, and test partitions, roughly consisting of 80\%, 10\%, and 10\% of the overall number of records, respectively. For the Random Uniform 1000 m/s RV Spectra, the train, validation, and test splits are composed of 1657680, 207180, and 207300 spectra, respectively. The validation split is consistently referenced during the training process to ensure the network is not overfitting. It further acts as a way to measure iterative changes made to the network design and their effect on the network’s performance. The held-out test data is only used for reporting the network’s performance and has no influence on the network development.

Each split is randomly sampled by referencing the unique timestamps associated with the HARPS-N spectra. This was done intentionally so that all spectra with a random RV shift applied to the same underlying spectrum can be collectively assigned to either the train, validation, or test split. Sampling the unique timestamps to create these splits also allowed us to generate the same splits based on timestamps for both the random RVs and the planet-like RVs, for training the neural network and evaluating the network's planet detection capabilities, respectively.

\subsection{Feature Scaling}\label{sec:data_prep_scaling}
Scaling of the input spectra and the target output RVs is performed to prevent large error gradient values and attain faster convergence in a gradient-based learning process, as is the case with neural networks. Spectra were zero-centered by subtracting the mean flux from all spectra, while the RVs were min-max scaled remapping the values to a range of [0, 1]. These specific strategies were selected by experimenting with different scaling approaches and selecting the scalers that resulted in the best performance. Scalers were fit using the training data and applied to rescale the validation and test data.

\subsection{Outlier Removal}\label{sec:data_prep_outliers}
Throughout the development of the neural network, a series of observations repeatedly presented themselves as problematic, having large errors relative to the collective distribution of errors. In total, 44 spectra from the original HARPS-N data were excluded for this reason. This list is available upon request from a corresponding author.


\section{Machine Learning Architectures} \label{sec:ml}
We developed a series of neural network-based approaches to isolate the planet-induced RV signal. We focused on designing network architectures that had significantly different characteristics regarding how spectra were handled. For each of the three approaches, the generic input to the neural network are the RV shifted spectra, while the target outputs are the RVs. These approaches result in a functional mapping where, given an input spectrum, the neural network will output an estimated RV value for a planet or planets.

\subsection{Single Line CNNs}\label{sec:ml_single_line_cnns}
A single CNN architecture was used to train networks for each spectral line. From this approach, we end up with 4570 CNNs - one CNN per spectral line - where each network is able to generate a predicted RV value for a given spectrum. 

The underlying CNN architecture used across all spectral lines consisted of 4 1-D convolutional layers (filter size of 3; step size of 1), followed by 3 fully connected dense layers. When the feature map output by the final convolutional layer is flattened, we concatenate feature map with the original 15 pixel spectral line input, all of which gets fed to the dense layers. 

For a given CNN, the RMSE for the predicted RV is calculated on the validation split, providing a single performance metric associated with that CNN. By scoring all 4570 line-based CNNs, we are able to rank the CNNs by RMSE. This ranking gives us a way to order lines based on their predictive capability for identifying (injected) planetary RVs.

\begin{figure}
\includegraphics[width=\columnwidth]{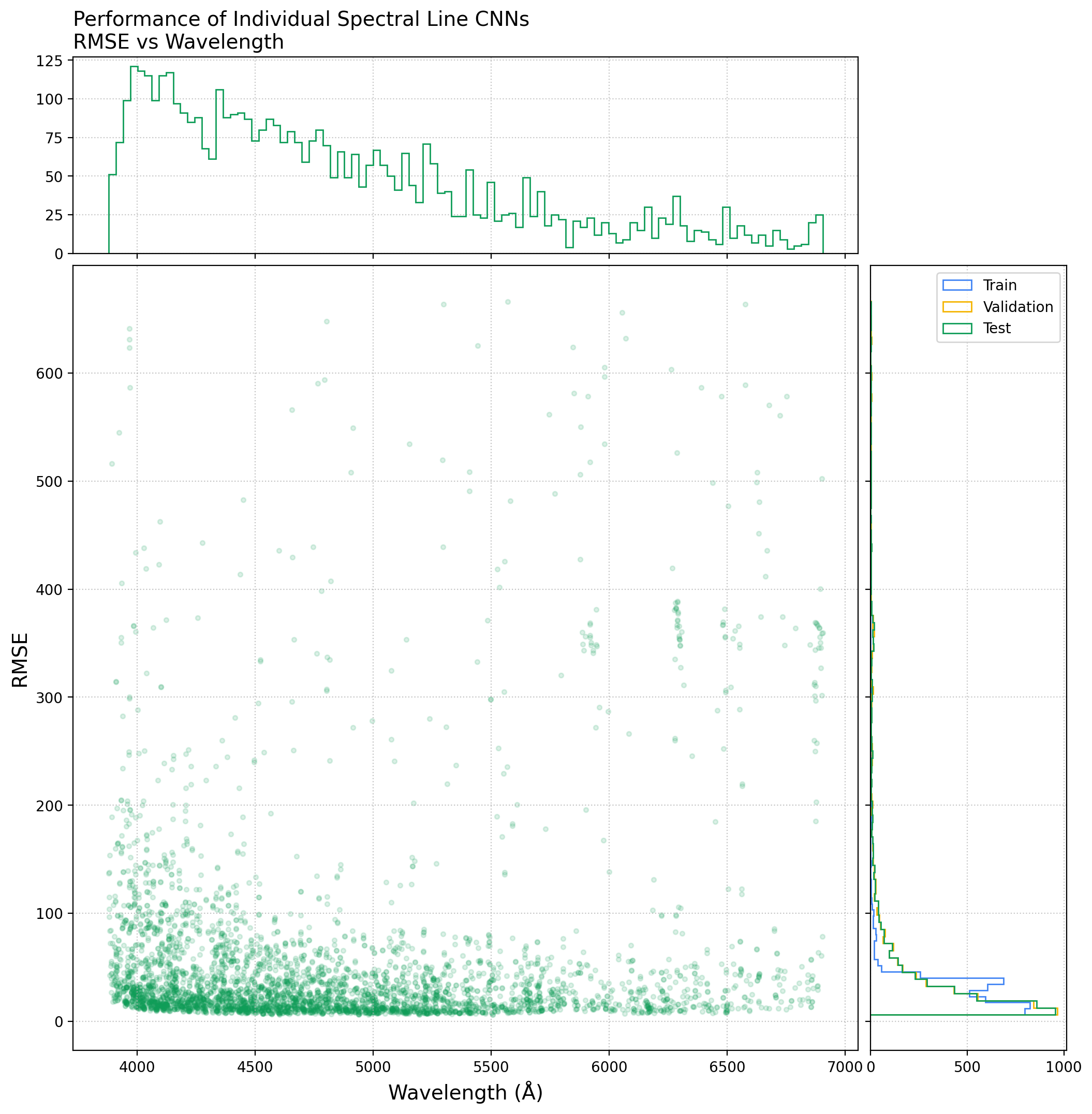}
\caption{RMSE vs wavelength for the center of each spectral line. The RMSE in the scatter plot is calculated on the test split. Histograms are provided for each axes to show: 1) how spectral lines are distributed across wavelength given the line list we used; 2) performance distribution (RMSE) for the train, validation, and test splits.}
\label{fig:scatter_n_rvs}
\end{figure}

The individual CNNs are not particularly useful given the best performing single-line CNN has a RMSE of 6.0183 m/s on the test set (Table \ref{tab:ml_results}, first row). However, the CNNs' RMSE scores have a positively skewed distribution (Figure \ref{fig:scatter_n_rvs}) indicating that the majority of lines offer some utility in measuring RVs. 

There are two follow-on approaches that take advantage of these networks and their outputs. One approach uses the predictions from the individual CNNs as inputs to train an ensemble, while the second approach takes a subset of lines with the lowest RMSE and collectively inputs those lines into a single CNN.

\subsection{Ensemble of Single Line CNNs}\label{sec:ml_ensemble}

\begin{figure}
\includegraphics[width=\columnwidth]{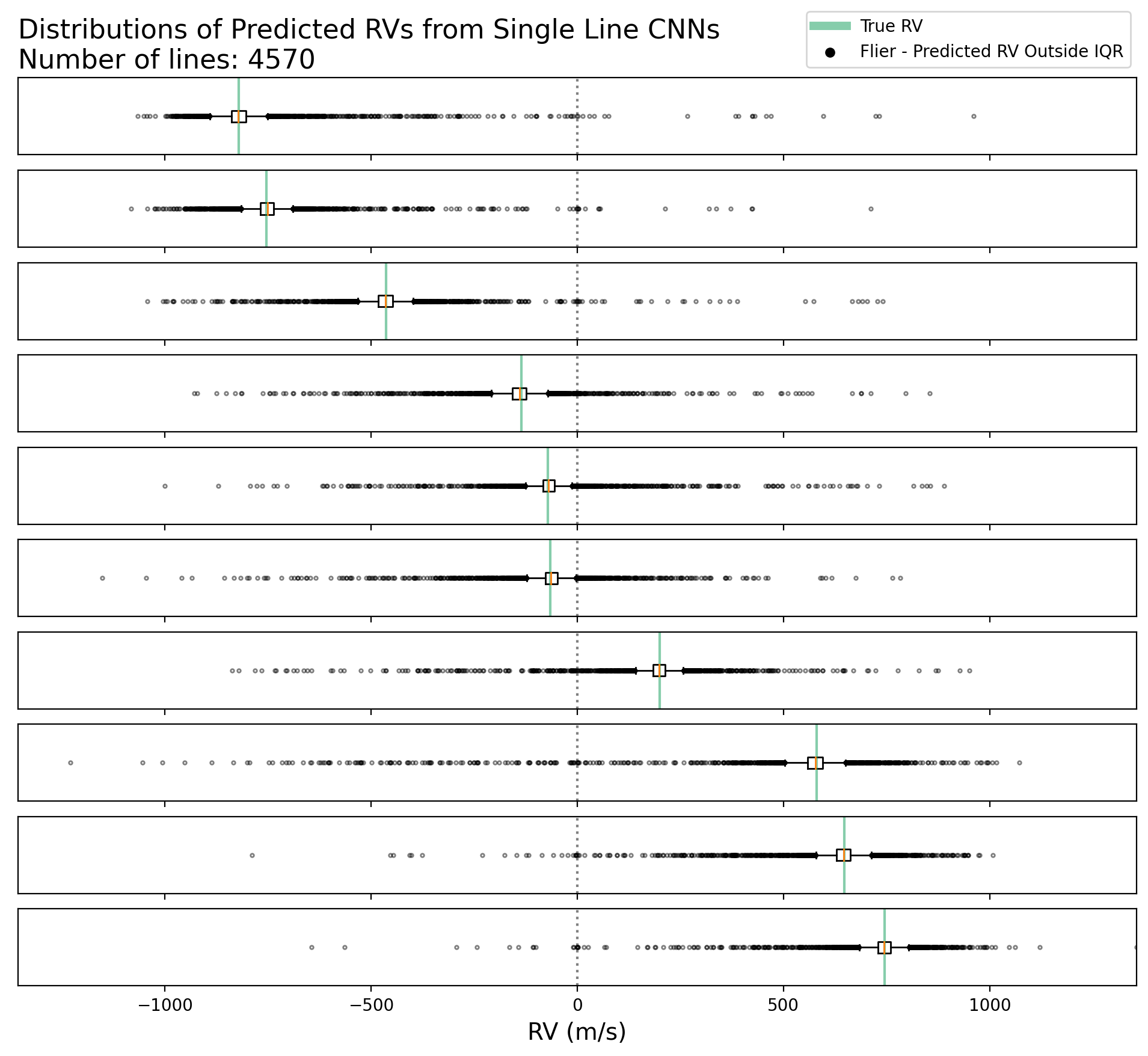}
\caption{Distributions of predicted RVs from the 4570 single line CNNs for a sample of the injected RVs. The distributions for each set of predictions is shown as a boxplot - red vertical line is the median, box is the IQR, whiskers are [Q1 - 1.5 * IQR, Q3 + 1.5 * IQR], and the black points any predictions outside the whiskers.  The vertical green lines correspond to the true RV values. The distribution for each set of predicted RVs is centered around the true RV (green vertical line).}
\label{fig:ensemble_inputs}
\end{figure}

\begin{figure}
\includegraphics[width=\columnwidth]{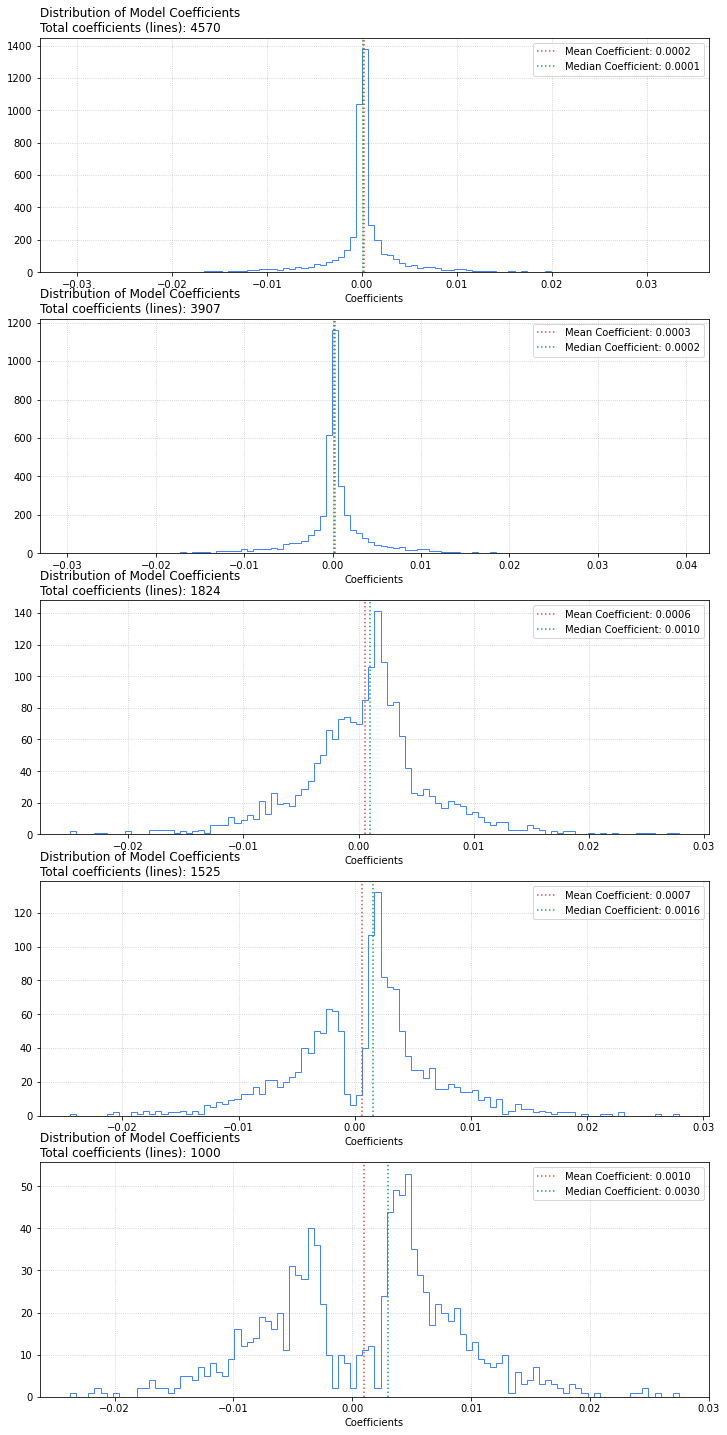}
\caption{Distributions of coefficients for the fitted OLS model using the single-line CNNs' predictions as inputs. The top distribution corresponds to the coefficients of the OLS model fit using all 4570 single-line CNNs' predictions. The subsequent distributions, from top to bottom, are the resulting distributions of fitted coefficients following pruning, where any input with a coefficient approximately close to zero (LT 1e-4) is removed. }
\label{fig:ensemble_coefs}
\end{figure}


Ensembles are useful in that they learn to recognize the strengths and weaknesses of multiple models in conjunction with one another. The model that combines the predictions from other models, CNNs in this case, is referred to as a meta-learner. A common approach is to use an ordinary least squares (OLS) regression model to server as the meta-learner. Ensembles provide benefits over a single model in that ensembles are not fully reliant on a single set of weights and inputs, thus allowing them to better generalize to a given task.

Given the approach in \ref{sec:ml_single_line_cnns}, we combine the output RV predictions for each single line CNN. For the Random Uniform 1000 m/s RV Spectra, this results in a 2073000x4570 matrix of predicted RVs. Figure \ref{fig:ensemble_inputs} provides a sample of the inputs used to fit the OLS model. The OLS model is fit using the subset of predicted RVs that corresponds to the training split used to train each single line CNN.  

The ensemble OLS model was initially fit using the CNNs' predictions from all 4570 spectral lines, resulting in a test RMSE of 2.0939 m/s (Table \ref{tab:ml_results}). Examining the distribution of the model coefficients showed that many of the coeffincients were approximately zero (Figure \ref{fig:ensemble_coefs}). By iteratively pruning inputs associated with fitted coefficients approximately close to zero (less than 1e-4), and refitting the OLS model with the reduced number of inputs, the input dimension for the OLS model was reduced to 1000 variables (single line CNN RV predictions). The OLS model fit with the reduced input had a test RMSE of 1.2673 m/s (Table \ref{tab:ml_results}). A final meta-learner was fit after additional pruning, resulting in inputs consisting of 250 RV predictions. This model generalizes better than the ensemble with all inputs, but performs overall worse than the 1000 input ensemble.

A downside to this approach occurs when evaluating other datasets. In order to generate RV predictions for a planet-like RV dataset, all 1000 CNNs (associated with the pruned ensemble model) need to be loaded and individually used to generate predictions that feed into the ensemble OLS model.

\subsection{Multi-line CNN}\label{sec:ml_multi_line_cnn}

\begin{figure*}
\includegraphics[width=\textwidth]{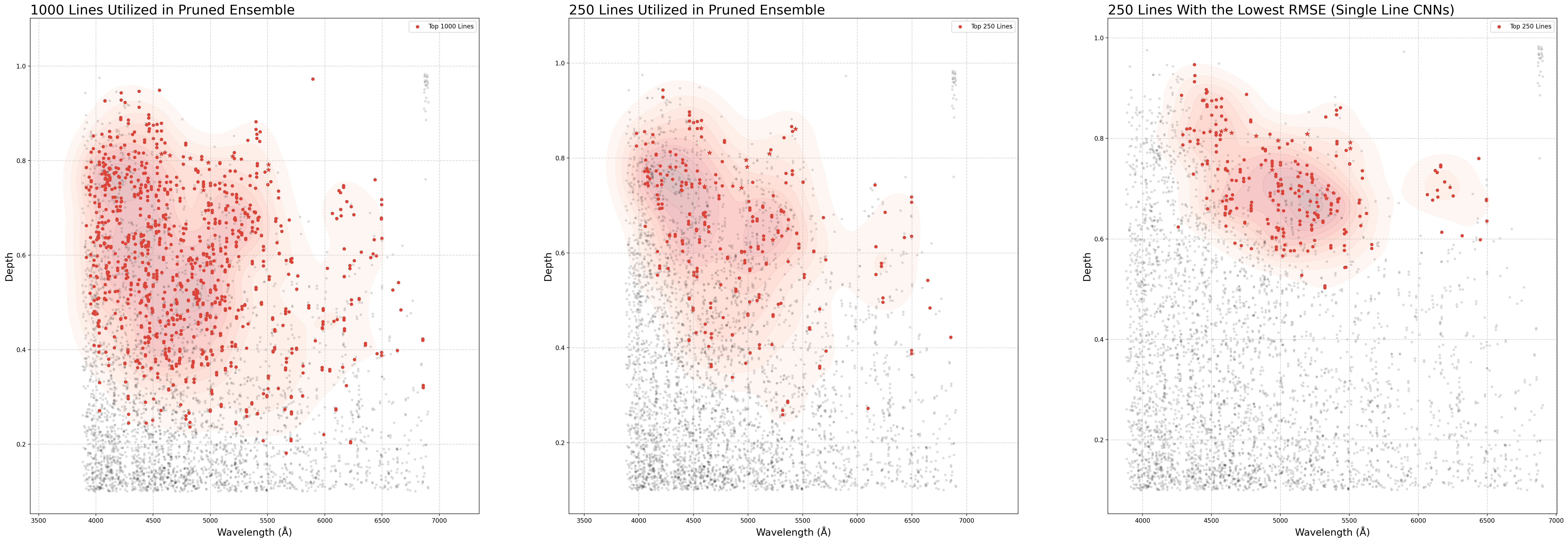}
\caption{Line depth vs line wavelength plots for the 4570 spectral lines used throughout this analysis. The left plot shows the 1000 lines, colored in red, resulting from pruning in Section \ref{sec:ml_ensemble}, while the center plot shows the 250 lines resulting from further pruning. The right plot shows the 250 spectral lines with the lowest test RMSE from the single line CNNs. The multi-line CNN was trained once using the lines in the center plot and a separate time using the lines in the left plot. }
\label{fig:multi_cnn_depth_v_wave}
\end{figure*}

To address the complexity of orchestrating hundreds of CNNs, we took an alternative approach that utilizes a single CNN with multiple spectral lines as inputs. 250 spectral lines were stacked such that each line is represented in a channel or band-like structure, similar to the RGB channels of a color image. A single input to the muti-line CNN is a 250x1x15 dimensional array where each vector indexed along the 0th axis corresponds to a single spectral line. The number of lines, 250, was selected based upon our remote compute environment's system memory and GPU memory balanced against the size of the Random Uniform 1000 m/s RV dataset.


 We considered two different sets of 250 spectral lines while training the CNN. The first set of spectral lines consisted of the 250 single line CNNs with the lowest RMSE (right plot, Figure \ref{fig:multi_cnn_depth_v_wave}). The second set of 250 spectral lines were derived from additional pruning of the 1000 spectral line ensemble (center plot, Figure \ref{fig:multi_cnn_depth_v_wave}). 

The architecture of this multi-line CNN is the same as the architecture used for the single line CNNs, with the exception that the input layer is changed to accommodate 250 spectral lines, opposed to a single spectral line. In this CNN, the weights of the filters in the convolutional layers are learned feature extractors where the features are learned from  all the input spectral lines collectively. This differs slightly from the single line CNN, where the weights of the filters are feature extractors that are learned from and specific to a given line.

Overall, the multi-line CNN trained on the 250 best individual lines seems well balanced, where the spread of RMSE scores across the train, validation, and test splits are not that dissimilar from one another (Table \ref{tab:ml_results}, CNN (Top 250)). This network performs slightly better than the 1000 spectral line ensemble (Table \ref{tab:ml_results}, Ensemble (1000)). The multi-line CNN trained on the 250 lines remaining from further pruning of the 1000 line ensemble, achieves the best RMSE on the training data, but has issues generalizing to the validation and test data (Table \ref{tab:ml_results}, CNN (250 Pruned)). 

\begin{figure}
\includegraphics[width=\columnwidth]{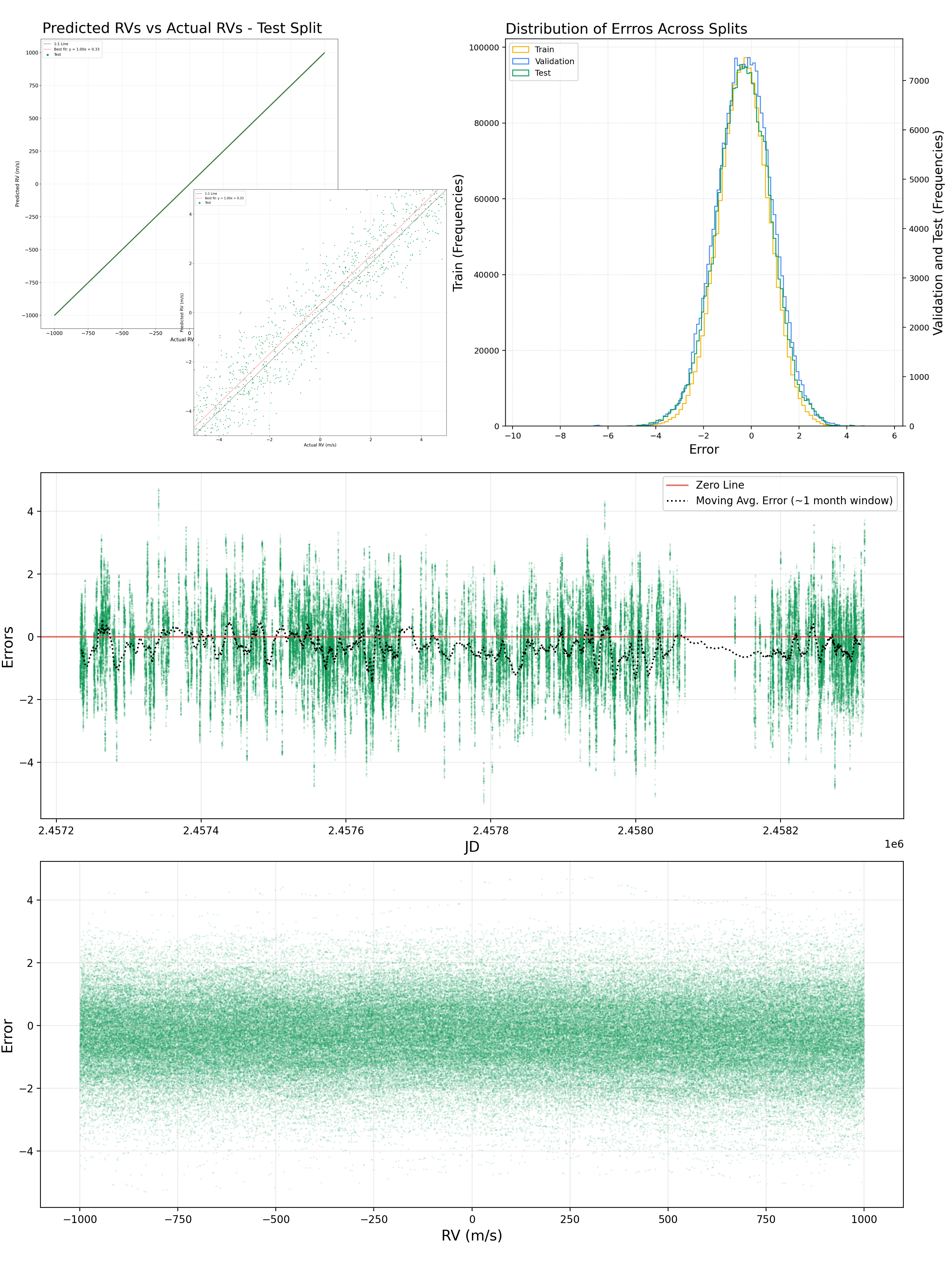}
\caption{A series of error analysis plots used to assess systemic issues associated with a trained neural network. Predicted RVs and the associated errors are from the multi-line CNN (250 Top). The displayed plots are from test data, however, in the development of the neural network the training and validation splits were referenced for informing necessary changes to in preprocessing steps or network architecture design; test data is show only for reporting in this context.}
\label{fig:multi_cnn_errors}
\end{figure}

Figure \ref{fig:multi_cnn_errors} outlines a series of error (residual) analysis plots. The plots pertaining to the test split for reporting in this document, but in practice utilized the validation split for assessing systemic issues present in the errors. Noteworthy aspects of these plots include: a near perfect linear relationship between the predicted RVs and actual RVs (upper left, Figure \ref{fig:multi_cnn_errors}); normally distributed errors (upper right, Figure \ref{fig:multi_cnn_errors}) albeit not zero centered; no distinct temporal trends (middle row, Figure \ref{fig:multi_cnn_errors}); and no apparent trends or patterns in the full range of RVs or a localized subset of RVs (bottom row, Figure \ref{fig:multi_cnn_errors}).

\begin{table}
\begin{tabular}{lrrr}
\multicolumn{4}{r}{RMSE (m/s)} \\
\cline{2-4}
Architectures & Train & Validation & Test \\
\hline
Single Line CNN   &  6.1177  &  6.0839  &  6.0183  \\
Ensemble (4570)   &  1.0360  &  2.1124  &  2.0939  \\
Ensemble (1000)   &  1.1964  &  1.2540  &  1.2673  \\
Ensemble (250)    &  1.4027  &  1.4502  &  1.4448  \\
CNN (Top 250)     &  1.1051  &  1.2213  &  1.2069  \\
CNN (Pruned 250)  &  1.0282  &  1.3729  &  1.3430  \\
\hline
\end{tabular}

\begin{tabular}{lrrr}
\multicolumn{4}{r}{MAE (m/s)} \\
\cline{2-4}
  & Train & Validation & Test \\
\hline
Single Line CNN   &  4.8340  &  4.8463  &  4.7543  \\
Ensemble (4570)   &  0.8173  &  1.7002  &  1.6756  \\ 
Ensemble (1000)   &  0.9427  &  0.9921  &  0.9949  \\
Ensemble (250)    &  1.1061  &  1.1413  &  1.1415  \\
CNN (Top 250)     &  0.8756  &  0.9693  &  0.9530  \\
CNN (Pruned 250)  &  0.8269  &  1.0890  &  1.0599 \\
\hline
\end{tabular}
\caption{Performance (RMSE and MAE) of different architectures on the train, validation, and test data. Performance for the ensemble models includes the full input data corresponding to all 4570 lines, 1000 lines after pruning, and 250 lines after additional pruning.}
\label{tab:ml_results}
\end{table}

\section{Results} \label{sec:results}

Our approach utilizes NNs to filter the RV contribution from stellar activity in a given input spectra by explicitly forcing the network’s attention on what we care about the most - the (injected) planetary RVs - during the training process. As such, the predicted RVs are, in the best case scenario, exact representations of the true planetary RVs, however, the trained neural network is imperfect where the predicted RVs contain error. The output estimated planetary RV signal is still useful. 

To better understand the extent of the neural network's utility in finding planets, we use the multi-line CNN (250 lines) to make RV predictions for each of the planet-like RV datasets described in Section \ref{sec:data_prep_rvs_planet}. Because these planet-like datasets are created using the same underlying HARPS-N solar spectra, which was also used to create the 1000 m/s Random RV dataset, we use the same spectra in the test split, used to score the neural network, to evaluate and check the network's utility in detecting planets. 

Despite training on a large quantity of diverse spectra, the neural network trained on the 1000 m/s Random RV dataset still has a small vertical offset in the predicted RVs when compared to the actual RVs for the training data. Hence, for each planet-like RV datasets' predicted RVs, we apply a standard correction corresponding to the intercept coefficient from the best fit line for the predicted vs actual RVs (upper left, Figure \ref{fig:multi_cnn_errors}) for the training data. Since planetary mass measurements do not require an accurate aboslute RV scale, but rely only on differential RV measurements, this offset correction is not considered to be a weakness in our approach.  

\begin{figure*}
\includegraphics[width=\textwidth]{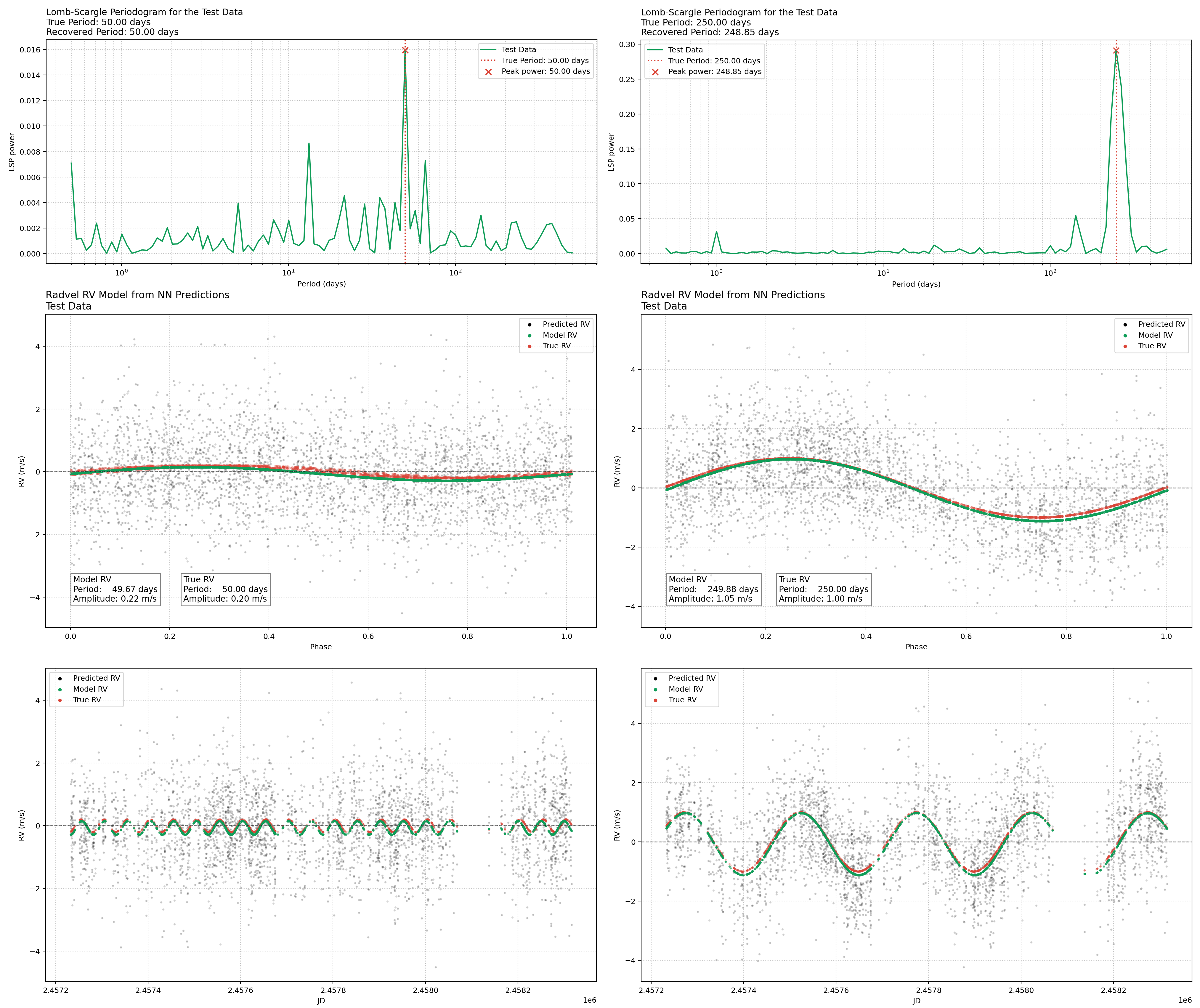}
\caption{Two sets of results for the planet-like test cases. The left column corresponds to a planet with a 50 day period and 0.2 m/s semi-amplitude, while the right column corresponds to a planet with a 250 day period and 1.0 m/s semi-amplitude. The top row depicts power and frequency for the Lomb-Scargle periodograms fit to each planet-like test case. The red dotted line is the true period, while the red "x" indicates the frequency of the peak power. The center row shows phase-folded diagrams while the bottom row shows the time series plots, both produced following the MLE. Dark points in the phase-folded diagrams and time series plots are the predicted RVs made by the neural network - predicted RVs for the 1.0 m/s semi-amplitude planet show distinct sinesoidal characteristics.}
\label{fig:multi_cnn_amplitude_period_test_results}
\end{figure*}

Following this correction, we fit a Lomb-Scargle periodogram to a given planet-like test case's predicted RVs. We take the period associated with the maximum power of the periodogram to correspond to the recovered orbital period of the injected planet (top row, Figure \ref{fig:multi_cnn_amplitude_period_test_results}). We assume, given a neural network capable of perfectly predicting a planet's RVs, we should be able to recover any period, constrained to the sampling window and cadence.

Using RadVel \citep{fulton18}, we perform a maximum likelihood estimation (MLE) fit to recover the amplitude of the injected planet. For initial parameter guesses, we set: the period to the highest peak in the Lomb-Scargle periodogram; the semi-amplitude to the 90th quantile of the predicted RVs; RV jitter to the training RMSE score from the 1000 m/s Random RV dataset. After fitting, we compare the model estimates for period and amplitude against the true period and amplitude. 

\begin{figure*}
\includegraphics[width=\textwidth]{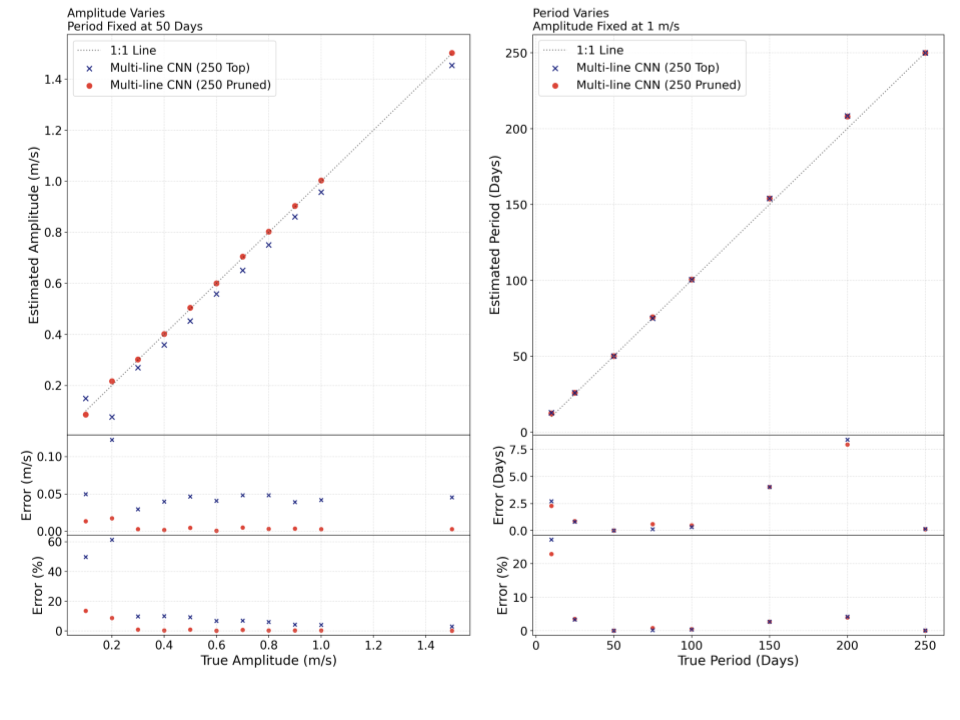}
\caption{Performance results for the 20 planet test cases. The left column corresponds to the amplitude recovery test cases and the right column corresponds to the period recovery test cases. Purple "x"s correspond to the multi-line CNN (250 Top) RVs and the red circles correspond to the multi-line CNN (250 Pruned) RVs. While both networks performed similarly on the period recovery test cases, the multi-line CNN (250 Pruned) outperformed the multi-line CNN (250 Top) on all amplitude test cases in terms of absolute error (middle left) and percent error (bottom left).}
\label{fig:multi_cnn_test_results}
\end{figure*}

The complete results for this process, performed on all 20 injected planet test cases, are captured in Figure \ref{fig:multi_cnn_test_results}. For this analysis, we produced results with both multi-line CNNs referenced in Section \ref{sec:ml_multi_line_cnn}, "250 Top" and "250 Pruned". While the 250 Top CNN was more balanced in regards to its train, validation, and test RMSE scores on the 1000 m/s Random RV dataset, we saw it underperform on the task of recovering the planet-like RV signals from the injected planet test cases. In comparison, the 250 Pruned CNN performed well when recovering the planet-like RVs, even on the spectra associated with the test split it had struggle to generalize as well to. In the amplitude sensitivity tests, we were able to recover a 0.2 m/s semi-amplitude, 50 day period, planetary signal, where the periodogram and MLE fit identified the true amplitude and period, within 8.8\% error and 0.7\% error, respectively (left column, Figure \ref{fig:multi_cnn_test_results}). In every amplitude test case, the 250 Pruned CNN outperformed the 250 Top CNN. Both CNNs had nearly identical performances on the period test cases, where they struggled to identify a planet with an amplitude of 1.0 m/s and period of 10 days, but succeeded on every planet with an amplitude of 1.0 m/s and a period of 25 days or greater. 

\begin{figure*}
\hspace*{-2cm}
\includegraphics[width=1.25\textwidth]{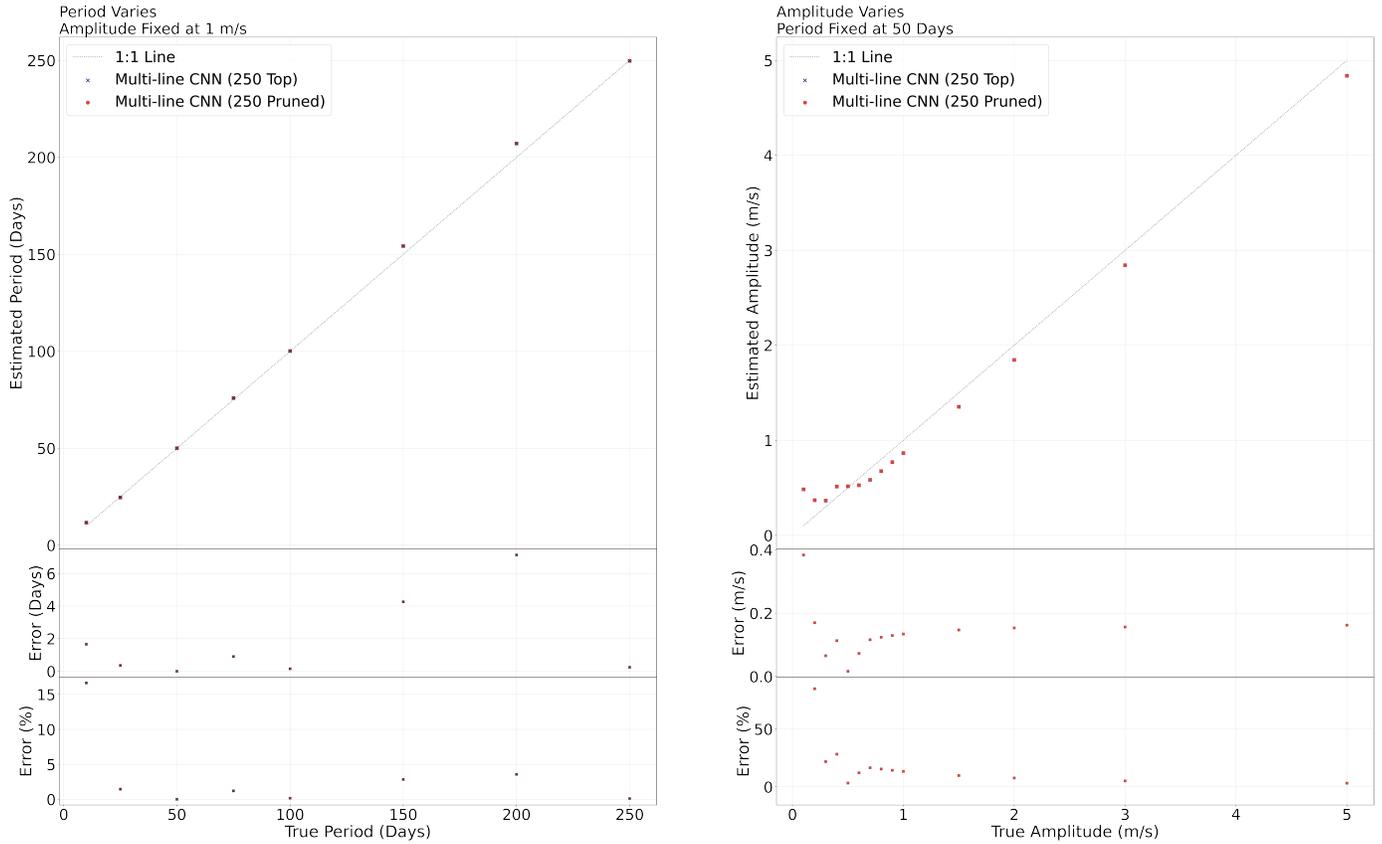}
\caption{Performance results for the 20 planet test cases using HARPS-N RVs in the solar frame. The left column corresponds to the period recovery test cases and the right column corresponds to the amplitude recovery test cases {\bf (note this is switched from the previous figure)}. This figure may be compared with Figure \ref{fig:multi_cnn_test_results} to assess how much our approach improved planet detectability. We see a substantial reduction in error in recovered planet parameters in our multi-line CNN approach when compared with these results, where we apply the same planet-fitting method to ``uncorrected" RVs.}
\label{fig:multi_cnn_test_results_harpn}
\end{figure*}

To assess how much our approach was able to mitigate noise in the RVs and improve planet detectability, we generated another set of ``results" by simply adding the sinusoidal planet RV signals to the HARPS-N RVs in the solar frame. The represent an uncorrected case, where due the large quantity of data ($\sim$3500 points over 3 years in our test set) we are still able to fit our planet model with some success. The results of this comparison study are shown in Figure~\ref{fig:multi_cnn_test_results_harpn}. {\bf We apologize for the formatting differences, these will be corrected in a future draft.} As a comparison to our CNN approach, where we recovered a 0.2 m/s semi-amplitude planet to 8.8\% error, here the same lomb-scargle periodgram + RadVel model approach recovered this 0.2 m/s signal with 80\% error, about an order of magnitude higher. In this experiment, the minimum planetary signal that could be added to the CCF RVs while maintaining a 10\% error threshold was 50 cm/s. As a result of this comparison, we suggest that our multi-line CNN approach for measuring RVs has the potential to substantially improve RV measurement precision once it has been developed to apply to non-solar targets.

\section{Conclusion and Discussion} \label{sec:conclusiondiscussion}

Using the public release of HARPS-N solar spectra, we demonstrated a deep learning based approach to recover injected planetary RVs in spectra. This approach is unique in that it operates specifically in the wavelength domain and does not utilize temporal features - lagged observations or aggregated observations. The neural network is trained on a diverse collection of spectra, injected with random RVs ranging from [-1000.0, 1000.0] m/s, which allow the network to generalize well to arbitrary sinusoidal planets. We apply the trained neural network to complete sinusoidal planets and utilize a Lomb-Scargle periodogram on the network's predicted RVs, showing that the frequency tied to the planet's period is associated with the highest power, down to a 10 day period and 1 m/s semi-amplitude planet. Additionally, we show that a planet with a 50 day period and 0.2 m/s semi-amplitude is recoverable from the predicted RVs using MLE.  

Future work can include making improvements to the neural network's performance and our interpretation of the results. Specifically, this may include experimenting with different sampling techniques for train, validation, and test splits, improving selection of the line subsets to feed to the neural network and refining the training pipeline to operate beyond current memory constraints. Future work may also include more rigorous hyperparameter tuning across the entire pipeline and an overall improved reporting of results to include planetary mass and period uncertainties.

\acknowledgments
The research was carried out at the Jet Propulsion Laboratory, California Institute of Technology, under a contract with the National Aeronautics and Space Administration (80NM0018D0004). We thank Sam Halverson, Eric Ford, John Callas, Charles Lawrence, Jenn Burt for helpful ideas and suggestions.

\clearpage

\end{document}